\documentclass{article}

\usepackage[final]{styles/neurips_2024}
\usepackage{xspace} 
\usepackage[utf8]{inputenc} % allow utf-8 input
\usepackage[T1]{fontenc}    % use 8-bit T1 fonts
\usepackage{hyperref}       % hyperlinks
\usepackage{url}            % simple URL typesetting
\usepackage{booktabs}       % professional-quality tables
\usepackage{amsfonts}       % blackboard math symbols
\usepackage{nicefrac}       % compact symbols for 1/2, etc.
\usepackage{microtype}      % microtypography
\usepackage{xcolor}         % colors
\usepackage{graphicx}
\usepackage{wrapfig}

\newcommand{\method}{\textbf{\textsc{Racoon}}\xspace}
\newcommand{\etl}{\textit{\textsc{Entity-Labels}}\xspace}
\newcommand{\ett}{\textit{\textsc{Entity-Triplets}}\xspace}

\title{RACOON: An LLM-based Framework for Retrieval-Augmented Column Type Annotation with a Knowledge Graph}

\author{%
  Lindsey Linxi Wei, Guorui Xiao, Magdalena Balazinska \\
  University of Washington\\
  \texttt{\{linxiwei, grxiao, magda\}@cs.washington.edu} \\
}

\begin{document}

\maketitle

\begin{abstract}
  % With the ever-increasing numbers of Web Tables, the research community is putting more and more attention on data integration and data exploration to make full use of these tables. 
  As an important component of data exploration and integration, Column Type Annotation (CTA) aims to label columns of a table with one or more semantic types. With the recent development of Large Language Models (LLMs), researchers have started to explore the possibility of using LLMs for CTA, leveraging their strong zero-shot capabilities. In this paper, we improve on LLM-based methods for CTA by showing how to use a Knowledge Graph (KG) to augment the context information provided to the LLM. Our approach, called \method, combines both pre-trained parametric and non-parametric knowledge during generation to improve LLMs' performance on CTA.
  Our experiments show that \method achieves up to a 0.21 micro F-1 improvement compared against vanilla LLM inference.
  %\todo{since 0.21 for single-label, consider using multi-label best result instead?}
\end{abstract}

\section{Introduction}

%\textbf{Motivation.} 

In recent years, researchers and data scientists have found abundant relational tables on the Web (\cite{cafarella2008webtables, cafarella2008uncovering, bhagavatula2015tabel}). 
Despite their high quality, these Web Tables often miss important meta information such as column headers and relationships between columns (\cite{suhara2022annotating}).
This information is essential for downstream tasks such as data quality control (\cite{schelter2018automating}) and data discovery (\cite{chapman2020dataset}). 
A key step in recovering this meta information is \textit{Column Type Annotation} (CTA), which aims to assign one or more \textit{semantic type(s)} to columns in a given table. For example, the column ['UEFA Champions League', 'Scottish Cup'] has the annotation “time.event”.

Thus, there has been increasing interest and effort recently in developing methods that can \textit{automatically} assign semantic labels to columns (\cite{deng2022turl, suhara2022annotating, miao2024watchog}). 
Recent work (\cite{kayali2024chorus, feuer2023archetype}) argues that relying on \textit{Pre-trained Language Models} (PLMs) is not sufficient because they need task-specific and dataset-specific fine-tuning with \textit{carefully labeled training dataset} to achieve reasonable results.
On the other hand, the recent development of \textit{Large Language Models} (LLMs) (\cite{radford2018improving,brown2020language, ouyang2022training}) with pre-trained parametric memory has opened the possibility of bridging this gap by requiring minimal or even no training examples to achieve promising performance.

There has been much research effort in developing effective solutions that utilize LLMs for CTA and table related tasks (\cite{narayan2022can}, \cite{tian2024spreadsheetllm}). For CTA specifically, one line of work (\cite{korini2023column}) directly applies LLMs to solve CTA  with promising performance.
CHORUS (\cite{kayali2024chorus}) and Archetype (\cite{feuer2023archetype}) further develop other components using history and remapping to improve performances. 
Table-GPT (\cite{Li2024tablegpt}) \textit{fine-tunes} LLMs to adapt to relational data format to improve performance on various table understanding tasks. 

%Note that \method is an orthogonal work to the above existing works. More specifically, we observe that at the core of existing methods, they still directly apply LLMs \textit{without} any external knowledge while our work tries to \textit{retrieve} external knowledge from KG to augment prompt for LLMs.

At the core of existing methods, they directly apply LLMs without any external knowledge. Despite promising performance, LLM-based methods still encounter challenges dealing with outdated knowledge (\cite{he2022rethinking}), producing factual inaccuracies (\cite{ji2023survey}), and handling domain-specific or specialized queries (\cite{kandpal2023large}). To alleviate such problems, \textit{Retrieval-Augmented Generation} (RAG), which enhances LLMs' generation by retrieving knowledge from external sources as non-parametric memory (\cite{asai2023retrieval}), has emerged as a promising approach. Among the many external knowledge sources, \textit{Knowledge Graphs} (KGs), can provide succinct yet informative knowledge to help analyze the intricate semantics of entities (\cite{ji2021survey, wang2017knowledge}). Specifically, KGs represent information in the form of entities (nodes) and relations (edges) (\citet{wang2017knowledge}). Each edge is a factual triplet in the form of (subject, relation, object), making KG a well-structured data source.
% \lindsey{KGs represent information through entities (nodes) and relations (edges) (\citet{wang2017knowledge}). Each edge is a factual triplet in the form of (subject, relation, object), e.g., (Raccoon, color, gray).}
% KGs represent information through entities (nodes) and relations (edges) (\citet{wang2017knowledge}). Each edge is a factual triplet in the form of (subject, relation, object). 
% \guorui{should we discuss example? No space? e.g., (Raccoon, color, gray).}
%knowledge from KGs with effective yet efficient
Many recent approaches have studied integrating KGs with LLMs/PLMs through RAG to improve performances on various tasks (\cite{Xu2020ControllableSG, pan2024unifying, dehghan2024ewek, sun2024thinkongraph, wen2024mindmap}) and saw promising performance gains compared with vanilla LLM inference.
But none of the above approaches focuses on CTA. The most relevant work KGLink (~\cite{wang2024kglink}) incorporates an external KG with PLM \textit{during training} for CTA, thus encountering the exact PLM problems we mentioned before, whereas our work aims to leverage LLMs and KG to avoid expensive training or fine-tuning over large models.

\begin{wrapfigure}[16]{r}{0.3\textwidth}
    \centering    \includegraphics[width=\linewidth]{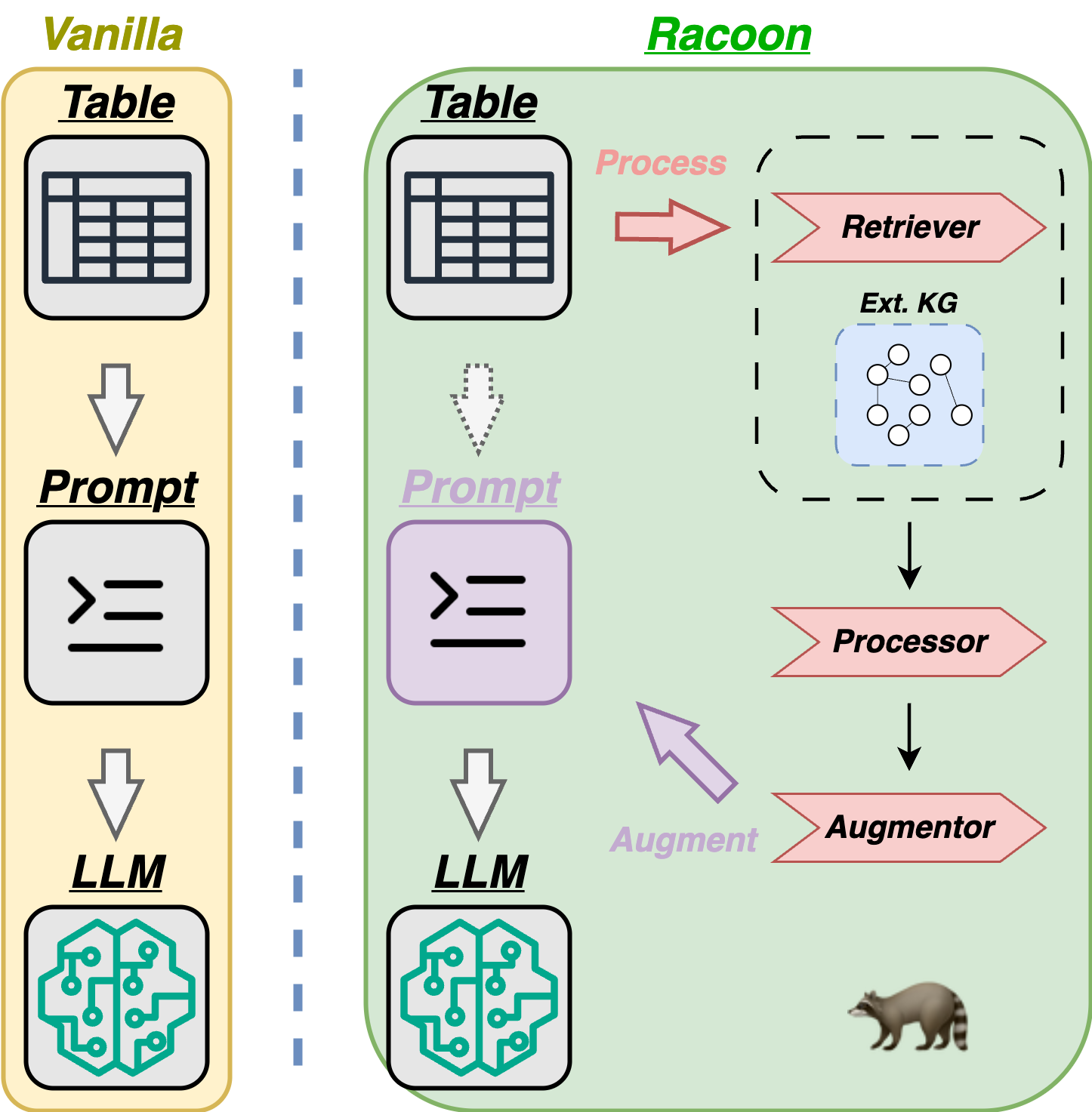} 
    \caption{Vanilla way of using LLM for CTA (left) vs \method (right).}
    \label{fig:racoon-intro}
\end{wrapfigure}

Inspired by these advances, we argue that we can similarly bring external non-parametric knowledge to LLMs for CTA to improve performance and propose our framework \method. As shown in Figure~\ref{fig:racoon-intro}, compared with vanilla LLM inference on the left, \method on the right unifies LLM inference with a KG by performing \textit{three additional steps}: namely after ingesting a table as the input, \method first (1) looks at the column cells' entity mention and \textbf{retrieves} related information from a KG. However, this retrieved content cannot be directly applied due to its large size and noise. Thus \method needs to (2) \textbf{process} the retrieved content, and finally (3) serialize the compressed context to \textbf{augment} the prompt for the LLM. Note that \method is orthogonal to existing LLM-based methods for CTA, which enhance LLMs' performance through additional demonstrations and response post-processing.

In summary, this paper makes the following contributions:

(1) We introduce the problem of augmenting prompts for LLMs on CTA with external knowledge from a KG and propose an end-to-end framework \method for it.

(2) We explore different granularity of information \method can retrieve from a KG with RAG. Furthermore, we provide effective post-retrieval compression and serialization methods such that given a column of cells, we can derive additional context information to augment the prompt.

(3) We conduct experiments showing that \method consistently outperforms vanilla LLM inference across various scenarios and retrieval methods, achieving up to a 0.21 improvement in micro F1.
% \magda{Repeat the main performance result here. This could be the 0.21 micro F1 improvementor another.}

%\textbf{Background.} 

% Specifically, it serializes the task alone with history of previous task and also develops a method called anchoring to help with CTA and other tasks such as join-column prediction. 
%Archetype (\cite{feuer2023archetype}), on the other hand, focuses only on CTA, but it develops other components such as context sampling and model querying to improve the usability and performance of LLM-for-CTA.

\section{Racoon's Approach}

\begin{figure}[h]
    \centering
    \includegraphics[width=0.9\linewidth]{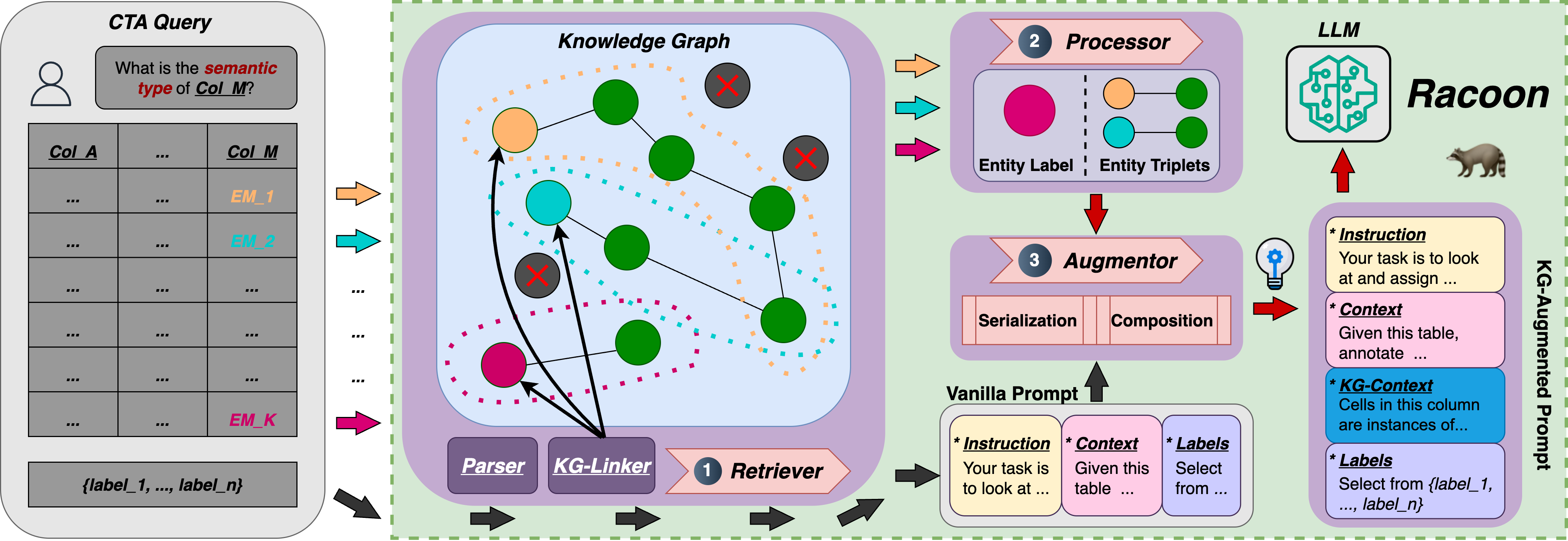}
    \caption{Overall \method workflow.
    \method performs CTA in a column-by-column manner while keeping the whole table for reference. The % \textbf{Retriever} finds the entities in the KG that are most related to the entity mentions in column cells and retrieves all relevant knowledge from KG.
    \textbf{Retriever} finds matching entities in the KG based on entity mentions in column cells and retrieves all relevant knowledge from KG.
    % and retrieves different levels of information from the KG. The %\textbf{Processor} reduces the retrieved information based on its granularity and statistics. 
    % \guorui{TODO}
    %\magda{Not clear what you mean by "based on its granularity and statistics? Can you find a better way to give the intuition here?} \guorui{How about the following? I wanted to hint just the purpose of each component here and explain the tech details below.} 
    \textbf{Processor} reduces the retrieved cell-level knowledge into an \textit{aggregated} column-level context. The compressed information is passed to \textbf{Augmentor}, which serializes all information into natural language and composes the KG-Context to form the final KG-Augmented Prompt.}
    \label{fig:racoon-main}
\end{figure}

We now describe the \method framework. The overall workflow is shown in Figure~\ref{fig:racoon-main}. 
Besides a CTA query, \method also requires an external KG and an LLM. Note that \method treats the KG and LLM as black-boxes and does not require any modifications, thus enjoying a plug-and-play property. %\edit{} 
For clarity of presentation, we describe how \method works with a CTA query on a single column, but \method can process CTA queries on an arbitrary number of columns.

\textbf{Retriever.} Inside the  Retriever, the \textit{Parser} first parses the input table based on the query. Given a table $T$ with a column $Col_M$ to perform single-column CTA, the Parser selects $C$ cells from $Col_M$ for retrieval purposes. As of now, \method uses all cells, and we leave the problem of selecting the most representative cells for future work.
The selected $C$ cells are treated as text strings and passed to the \textit{KG-Linker}, which takes entity mentions (the text strings) as input, connects with an external KG to link each entity mention to its referent entity in the KG, and returns the IDs of the referent entities. With those entity IDs, the final step is to further retrieve all entity labels and their one-hop neighbors in the KG as the output of the Retriever.
The task KG-Linker performs is defined as Entity Linking (EL), a popular task in table interpretation (~\cite{han2011collective, deng2022turl}). Since many EL methods are available online, \method allows users to implement their own KG-Linker based on the KG they decided to use. We have experimented with two KG-Linkers: a vanilla API and a SOTA EL model (~\cite{ayoola-etal-2022-refined}). Note that the Retriever assumes the KG-Linker can find entities in the connected KG. If it fails to link any selected cells, \method will fall back to using a vanilla prompt. 

% Right now, \method supports MediaWiki API as the retriever, but it allows user to define their own retriever by implementing a few virtual functions if necessary. 

%\todo{Talk about what challenges each component addresses?}

% \guorui{TODO}
%\magda{This is the most interesting paragraph in the paper. It would be great to find a way to expand it a bit to add more details: Why we tried both the ENTITY-LABELS and the ENTITY-TRIPLETS? Perhaps give an in-text example?}

\textbf{Processor.} The role of the Processor is to compress and refine the retrieved information from the KG, ensuring it is both relevant and concise. The retrieved information takes the form of sets of nodes and edges for each selected cell in $Col_M$, which is too large and noisy for LLMs to digest. Thus, the Processor takes them as input and outputs their concise representations. As of now, \method supports processing the retrieved information into two representations, \etl and \ett. 
\etl include labels of entities in the KG. The intuition is that these labels are canonical representations of entity mentions, which are often ambiguous or incomplete. For example, the entity mention `15 Sge' is linked to the entity `15\_Sagittae' in Wikidata KG, making it easier for LLMs to understand. Although \etl can disambiguate entity mentions, they do not provide any further information about the entities such as their types.
\ett, on the other hand, include triplets in the format (subject, relation, object). In our experiments, the linked entity serves as the subject, the relation is the \textit{instance\_of} relation, and the object is the entity type of which the subject is an instance. For example, (15\_Sagittae, instance\_of, star).
Note that \etl is strictly a subset of \ett. With this representation, \method fully leverages the ontological relationships among entities, enabling a deeper understanding of their types.
\method then uses statistics to count occurrences of each entity node and relationship triplet for all selected cells in $Col_M$ to give summary \textit{column-level} information, resembling the compress stage in traditional RAG systems. 
% \todo{refine statement here}

\begin{figure}
    \centering
    \includegraphics[width=0.75\linewidth]{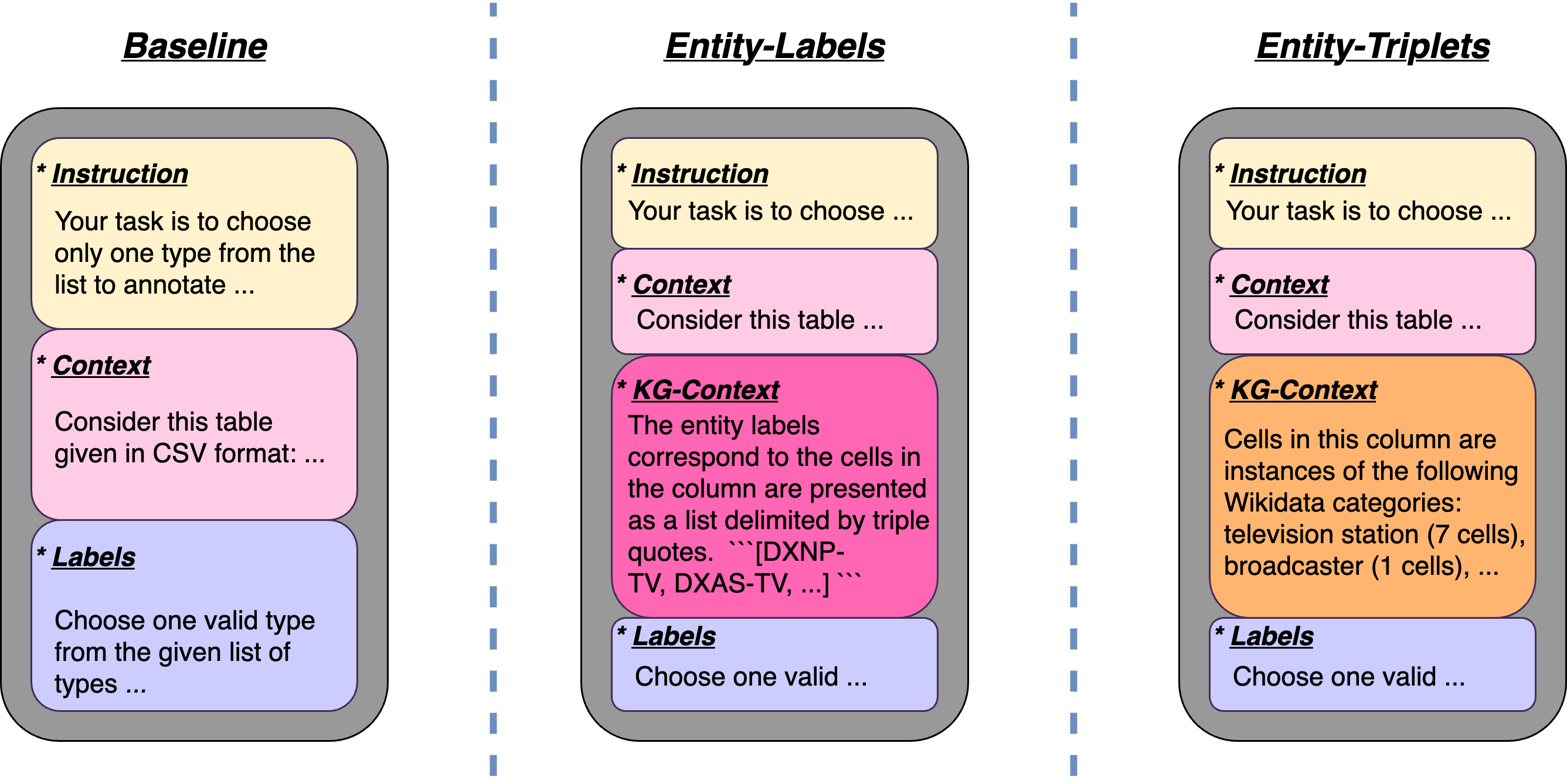}
    \caption{Vanilla (left) and \method (middle and right) prompts. \textit{Instruction} describes the CTA task in natural language, \textit{Context} describes the serialized table, and \textit{Labels} shows the label set. The middle prompt uses \etl and the right prompt uses \ett.}
    \label{fig:racoon-prompt}
\end{figure}

\textbf{Augmentor.} Finally, the Augmentor takes the original vanilla prompt and the processed representations from the Processor to create the KG-augmented prompt. It serializes the summarized column-level KG information into natural sentences and creates the KG-Context to insert into the original vanilla prompt. We show the detailed prompt in Figure~\ref{fig:racoon-prompt}. 
% However, if no information is given by the \textbf{Processor}, it will pass the vanilla prompt to the LLM directly.

\section{Experiments}
\textbf{Setup.} We evaluate \method using the GPT-3.5 model (~\citet{ouyang2022training}) and Wikidata KG (~\cite{vrandevcic2014wikidata}) on the full test set of the WikiTables-TURL-CTA benchmark  (~\cite{deng2022turl}) consisting of 13,025 columns extracted and annotated in a multi-label manner using 255 Freebase's types by the TURL team from the WikiTable corpus. The dataset provides a ground truth entity ID (Wikipedia page ID) linked to each cell in the table. 
On average, tables in this dataset have a mean of 21 rows and 4 entity columns (columns with at least one linked cell).

\textbf{Baselines.} We compare \method against a vanilla LLM method on CTA. Notably, both CHORUS and ArcheType (~\cite{kayali2024chorus, feuer2023archetype}) applies such vanilla LLM inference.
%CHORUS applied zero-shot prompting to the GPT-3.5 model, and we used the same model and inference prompt routine for our baseline experiment on CTA. \todo{clarify baseline}
 
% \textbf{Setup.} We use the GPT-3.5 model ~\citet{ouyang2022training} for our experiments. We generate responses with a temperature of 0 and a max token of 40. Our prompts have a token limit of 16,385 tokens, as determined by the context window of GPT-3.5.

\begin{table}
  \caption{The results of the baseline and \method in micro-F1 scores}
  \label{micro-f1-separate-columns-table}
  \centering
\begin{tabular}{l|c|ccc|cccc}
    \toprule
    &   & \multicolumn{3}{c|}{Multi-label} & \multicolumn{3}{c}{Single-label} \\
    \cmidrule(r){2-2} \cmidrule(r){3-5} \cmidrule(l){6-8}
    KG-Linker & EL & Baseline & ETL & ETT & Baseline & ETL & ETT \\
    \midrule
    Ground Truth  & 1.0000 & 0.2609   & 0.4023       & 0.4500    & 0.4631   & 0.564        & 0.6814  \\
    MediaWiki API  & 0.4018 & 0.2609   & 0.3582        & 0.3494   & 0.4631        & 0.4927            & 0.5222       \\
    ReFinED  &0.6255 & 0.2609        & 0.3698            & 0.3678    & 0.4631        & 0.4900            & 0.5705  \\
    \bottomrule  
\end{tabular}
\end{table}

\textbf{Evaluation.} We first use the ground truth entity ID linked to each cell labeled in the WikiTables-TURL-CTA dataset as the output of the KG-Linker to retrieve information from the KG. This enables us to evaluate our approach
assuming a perfect KG-Linker. The results in Table~\ref{micro-f1-separate-columns-table} show that both augmented prompts using \etl (ETL) and \ett (ETT) outperform vanilla LLM inference, with \ett achieving the best performance overall. This highlights the potential of retrieving more comprehensive information about column cells from the KG.

%The ground truth entity ID guarantees that the entity linked to each column cell is accurate, so we 

%the retrieved information is relevant and will positively impact \method's performance.
%\magda{Can you better explain the difference
%in the experiments between ground truth linking and the other two methods? This is important but not clear.} 
% \lindsey{The ground truth entity IDs guarantee that the retrieved cell-level information is correctly relevant to each cell. When the performance of the KG-Linker declines, incorrect entity IDs are linked to the cells, resulting in irrelevant information being appended to the prompt, which negatively impacts \method's overall performance.}

%\magda{I made a bunch of changes in the description here. Please check}
Second, we evaluate \method's performance with different KG-Linkers: MediaWiki API (with action=websearchentities) and a SOTA entity linking model, ReFinED (~\cite{ayoola-etal-2022-refined}). With either KG-linker, Racoon continues to outperform the baseline, although its performance (both \etl and \ett) drops compared with the ground-truth linker and the drop is higher for \ett, showing that
collecting extended information is not as helpful when the initial linked entity is incorrect. The drop is less with the ReFinED KG-linker, showing that SOTA KG-linkers suffice to make our approach beneficial.

Because LLMs often underperform in the multi-label dataset setting due to the large number of possible answer combinations,  we further test \method in the single-label setting: If the model prediction falls within the ground truth label set, we set the ground truth label to be the model prediction. Otherwise, we count it as an incorrect prediction. In this setting, the performance of both the baseline and \method improves. \method with ReFinED significantly outperform the baseline, with \ett giving the best results.

%outperforms the baseline significantly with \ett having a much better performance than \etl using ReFinED, further demonstrating the need for an effective KG-Linker.

% We also observed that the LLM sometimes provids reasonable answers, even if they don't exactly match the ground truth label. For instance, the LLM predicts `sports.sports\_team' while the ground truth label is `sports.sports\_league.' Both are among the 255 types in Freebase and have very close semantic meanings. 

\section{Conclusion}

We presented a novel end-to-end LLM-based framework \method for Column Type Annotation (CTA) unifying vanilla LLM and Knowledge Graph (KG) augmented inference. \method has three components: a Retriever for retrieving knowledge from a KG; a Processor for post-processing the retrieved content; and an Augmentor for composing the final prompt. Experiments show that \method \textit{consistently} outperforms vanilla LLM for CTA. 

%In the future, we plan to explore tailored and more effective Retriever and more dynamic Processor.
\clearpage
\section*{Acknowledgments} 
This project was funded in part by a gift from NEC.

\bibliographystyle{unsrtnat}
\bibliography{references/ref}
\clearpage
\appendix

\section{Appendix}
Here we show the example prompts for the single-label setting and the multi-label setting.
\subsection{Single-label}
\textbf{System message:}
Be a helpful, accurate assistant for data discovery and exploration desiged to output valid JSON in the format \{`type': []\}

\textbf{User message:}
Consider this table given in Comma-separated Values format:
                            ```
                            {Table}
                            ```
There are a list of 255 valid types for each column: {types}. Your task is to choose only one type from the list to annotate the first column. Solve this task by following these steps: 
1. Look at the cells in the first column of the above table. 
2. Consider this information carefully: Cells in this column are instances of the following wikidata entities: human (6 cells).
3. Choose only one valid type from the given list of types. Check that the type MUST be in the list. Give the answer in valid JSON format.

\subsection{Multi-label:}
\textbf{System message:}
Be a helpful, accurate assistant for data discovery and exploration desiged to output valid JSON in the format \{`type': []\}

\textbf{User message:}
For multi-label:
Consider this table given in Comma-separated Values format:
                            ```
                            {Table}
                            ```
There are a list of 255 valid types for each column: {types}. Your task is to choose one or multiple types from the list to annotate the first column. Solve this task by following these steps: 
1. Look at the cells in the first column of the above table. 
2. Consider this information carefully: Cells in this column are instances of the following wikidata entities: human (6 cells).
3. Mark each type in the given list with 0 or 1. Mark a type with 1 if it can better represent all cells of the first column. Mark a type with 0 otherwise. 
4. Give a list of types that you have marked 1 in the previous step. Check that the types MUST be in the list. Give the answer in valid JSON format.

\end{document}